\documentclass{aa} 
\usepackage{txfonts}
\usepackage{graphicx}
\usepackage{natbib}

\begin{document}
\title{A dam around the Water Fountain Nebula?}
\subtitle{The dust shell of IRAS16342-3814 spatially resolved with VISIR/VLT}

\titlerunning{}
    \author{
          T. Verhoelst\inst{1,2} \fnmsep \thanks{Postdoctoral Fellow of
the Fund for Scientific Research, Flanders}
          \and
          L.B.F.M. Waters\inst{1,4}
          \and 
          A. Verhoeff\inst{4}
          \and
          C. Dijkstra\inst{3}
          \and
          H. van Winckel\inst{1}
          \and
          J.W. Pel\inst{5}
          \and
          R.F. Peletier\inst{5}
          }

   \offprints{T. Verhoelst}

   \institute{            
    Insituut voor Sterrenkunde, K.U. Leuven, Celestijnenlaan
    200D, B-3001 Leuven, Belgium  \\         
    \email{tijl.verhoelst@ster.kuleuven.be}
    \and University of Manchester, Jodrell Bank Centre for Astrophysics, Manchester, M13 9PL, U.K. 
    \and Department of Physics and Astronomy, University of Missouri, Columbia, MO 65211, USA
    \and  Astronomical Institute ``Anton Pannekoek'', University of
    Amsterdam, Kruislaan 403, 1098 SJ Amsterdam, The Netherlands
    \and Kapteyn Astronomical Institute, Landleven 12, 9747 AD
    Groningen, The Netherlands 
}

   \date{Received; accepted }

   \abstract{
   Bipolar morphologies in Planetary Nebulae (PNe) are believed to be
closely linked to binary central stars. Either by collimating a fast
stellar wind or by driving a jet via accretion in the central system, dusty
torii or stable disks may be crucial ingredients for the shaping of
PNe.}  {We study the dust distribution in the very young
Proto-Planetary Nebule (PPN) IRAS16342-3814, also known as the Water Fountain
Nebula, which is known to show strong bipolar characteristics in the
shape of two reflection lobes, and high-velocity collimated molecular
outlfows.}  {We use the new Mid-IR (MIR) instrument VISIR on
the Very Large Telescope (VLT) both in imaging and spectroscopy mode
at wavelengths from 8 to 13\,$\mu$m.}  {We present the first spatially
resolved MIR observations of a dusty evolved star obtained with VISIR
and find that the improved spatial resolution contradicts previous
claims of an elliptical brightness distribution at the heart of
IRAS16342: we find the waist region to be dark even in the MIR.  We
show that the filling angle of the obscuring dust lane, which is made
mostly of amorphous silicates, is very large, possibly even close to a
spherically symmetric superwind as seen in OH/IR stars.}  {We conclude
that, in contrast to the multitude of recent dusty-disk detections in
Post-AGB stars and PNe, IRAS16342 does not show this extreme
equatorial density enhancement, at least not on the scale of the dusty
environment which lends the object its IR appearance. Rather, it
appears that the observed precessing jets are shaping the bipolar
nature in the remains of a spherically symmetric AGB superwind.}


   \maketitle
%

\section{Introduction}

Two different mechanisms for the shaping of bipolar and multipolar
(Proto-)Planetary Nebulae (PNe) are currently under intense
investigation \citep[e.g., ][]{APNIII}. The
first is in concept based on the Generalized Interacting Winds
model \citep{Balick1987} and suggests the collimation of the fast,
spherical wind of the hot central star by an older, slow, 
axi-symmetric and possibly warped, AGB (super-)wind
\citep[e.g.][]{Icke2003}.  The axi-symmetric shaping of the earlier
AGB wind may require the central star to be in a binary system.

The other scenario postulates that the lobes are in fact
cavities blown in a preceding spherical AGB wind by (precessing)
jets \citep{Sahai2005,Sahai2007}. The jets are then most likely the
consequence of (re-)accretion of circumstellar material, maybe also in
a binary system. This scenario is especially attractive to explain
very young PPNe, in which the central star is not yet hot enough to
exhibit a strong line-driven wind. Both hypotheses are not
mutually exclusive.

With the advent of powerful mid-IR interferometers, especially MIDI at
the VLTI, the type of dust disks which could collimate the fast
Post-AGB wind have now been observed around binary Post-AGB stars
\citep{Deroo2006,Deroo2007} and observations of the dust at the
heart of the Ant Nebula \citep{Chesneau2007} are consistent with a
disk interpretation.  On the other hand, optical and near-IR
observations of some PPNe show evidence for shaping by (precessing)
jets \citep{Sahai2007}.  The presence of a dusty torus is
also suggested in these objects, but it is hitherto unclear whether
these tori are similar to the disks observed with MIR interferometry.

IRAS16342-3814, hereafter IRAS16342 is a young PPN.  In spite of its extremely
red SED, with crystalline silicate features in absorption up to almost
45~$\mu$m \citep{Dijkstra2003}, optical HST images show a bipolar
reflection nebula with a dark equatorial waist \citep[][hereafter
STM99]{Sahai1999}.  According to STM99, the lobes are cavities blown
by a bipolar molecular jet, observed
as high-velocity lines of water (hence the designation ``water
fountain nebula'') and OH \citep{Likkel1988}, within the remains of a
{\sl low
mass-loss AGB wind which preceeded the recent superwind phase}.
\cite{Sahai2005} interpret local density enhancements in the bipolar
cavities as due to precession of the jet.  The dark waist, origin
of the bulk of the energy in the SED, is assumed to be an optically
thick dusty torus which completely obscures the central star in the
line-of-sight to the observer.

\cite{Dijkstra2003}, hereafter DKW03, present infrared ISAAC and
TIMMI2 images from 3.8 to 20~$\mu$m and find only at
the shortest wavelengths a bipolar nature such as in the HST
images. At longer wavelengths, the source appears elliptical, as was
already found by \cite{Meixner1999}. 

In this paper, we present a new
MIR image and an N-band spectrum of IRAS16342 obtained with the
VISIR instrument on the VLT, both of which resolve the dust
structure and force us to question our understanding of this (type of)
source.

\section{Observations}
\label{sec:observations}

The observations presented here were obtained as part of the Dutch GTO
on VISIR, the MIR imager and spectrometer installed at the Cassegrain
focus of Melipal at the VLT. An image was made in the SiC filter
($\lambda_c = 11.85 \mu m, \Delta\lambda = 2.34 \mu m$) with a
0.127~arcsec PFOV in the night of 20 March 2006, bracketed by 2 PSF
observations (HD~146051, M0.5III). Low resolution spectroscopic
observations in 4 bands (centered at 8.5, 9.8, 11.4 and 12.2~$\mu$m)
were made in the night of 17 March 2006 with a $32\times0.75$~arcsec
slit oriented 55.4$\degr$ East of North, and a 0.127~arcsec PFOV. The
calibrator was again HD~146051.  
science observations ranges from 1.03 to 1.05 and that 
standards from 1.08 to 1.12. The airmass of the imaging 
ranges from 1.15 to 1.25.  The visual seeing was about 0.60~arcsec for
the imaging observations and 0.90~arcsec at the time of the
spectroscopy.

For the data reduction, we use the pipeline offered by ESO, version
1.3.7. The synthetic N band spectrum of the calibrator is computed
using a {\sc marcs} atmosphere model \citep[][and further
updates]{Gustafsson1975}, including SiO opacity. The stellar
parameters are taken from \cite{Borde2002}.  The telluric correction
of the N band spectra is performed by solving for instrumental
efficiency and atmospheric optical depth using the 2 observations of
HD~146051. The resulting numbers are then interpolated in airmass to
match the conditions at the time of the science observation. As
absolute calibration, we scale the VISIR spectrum in such a way that
the simulated inband power over the IRAS 12~$\mu$m filter matches the
observed value of 16.2~Jy (see Sect.\,\ref{sec:spectra} for a
discussion on possible slit losses).

\section{Bipolar even in the Mid-IR}
\label{sec:imaging}

\begin{figure}
\centering
  \resizebox{7cm}{!}{\includegraphics{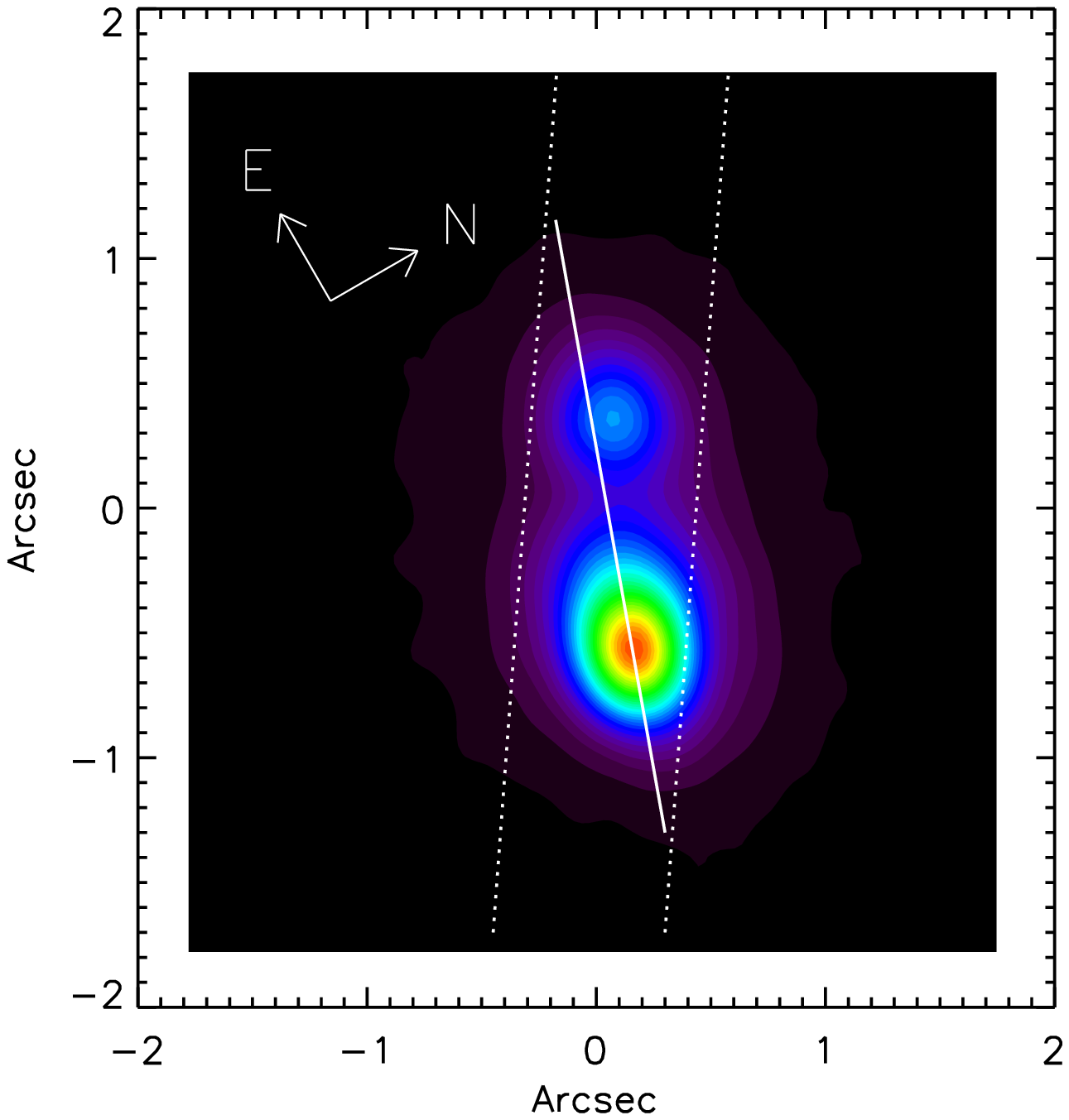}}
  \resizebox{7cm}{!}{\includegraphics{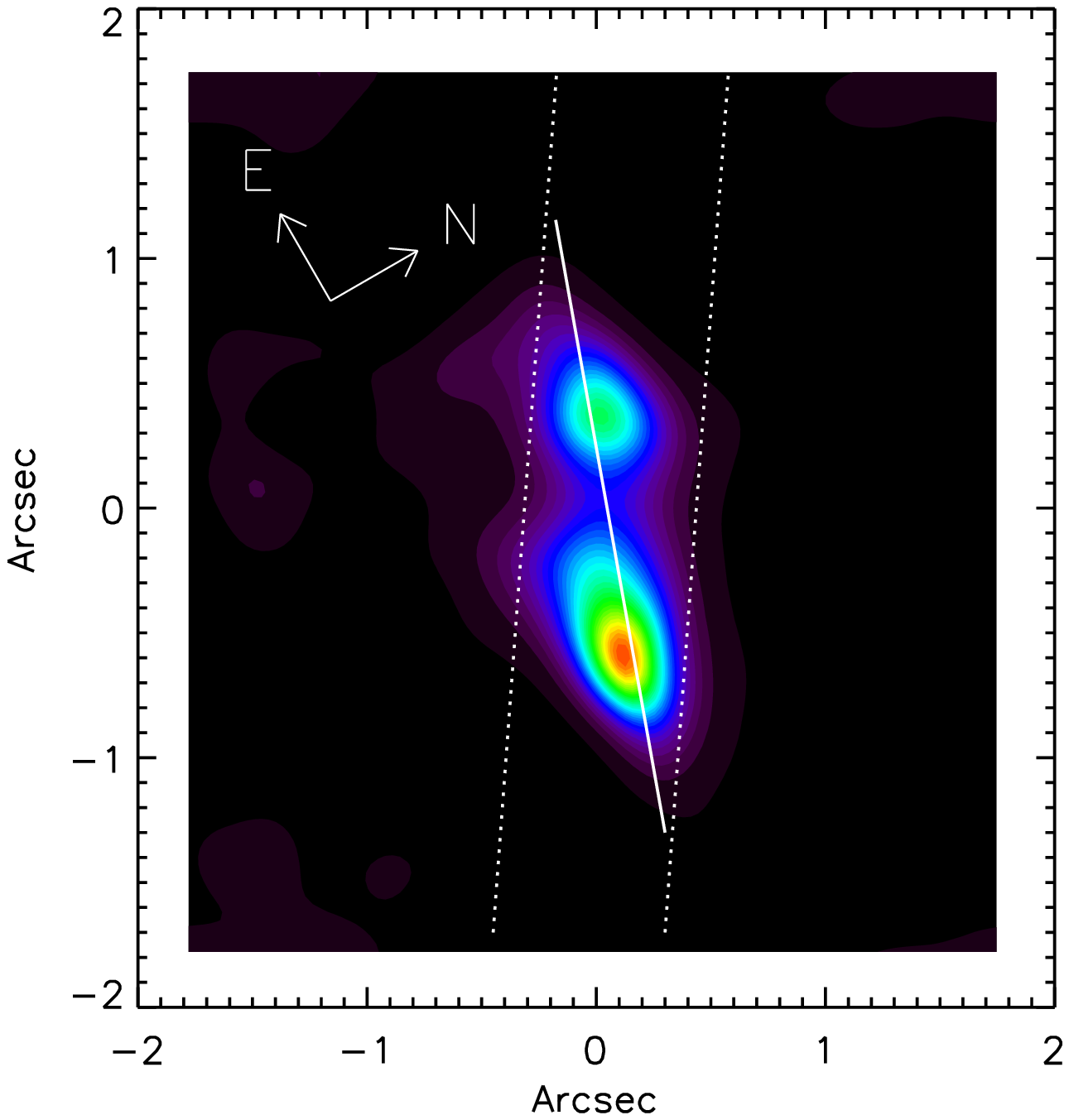}}
  \caption{
{\sl Upper panel:} The raw image of the Water Fountain Nebula in the SiC filter
 (11.85~$\mu$m). The colour scale is linear. The direction of the bipolar optical reflection
 nebula is also indicated (solid line, from STM99). The dotted lines
 represent the position and width of the slit used for the
 spectroscopic observations. {\sl Lower panel:} The same image but
 deconvolved using the PSF observation of HD\,146051 and with the
 colour scale following I$^{(1/2)}$. The solid and
 dotted lines have the same meaning as in the upper panel.
}
  \label{fig:sic_image}
\end{figure}

The image deconvolution is performed with both a maximum-likelihood
\citep{Richardson1972,Lucy1974} and a maximum-entropy method
\citep{Agmon1979}. Differences between the results of both methods are
minimal. The raw and deconvolved images are shown in Fig.~\ref{fig:sic_image}.

%
%

We observe a double-peaked intensity distribution with a separation of
0.92~arcsec and a Position Angle (PA) of 66$\degr$ East of
North. Component W (West) contains about 3/4 of the total flux,
component E (East) the remaining quarter. TIMMI2 images in the N and Q
bands observed by DKW03 do not resolve 2 separate emission peaks, but
show instead an elliptical shape with a major axis oriented more or less like
the separation vector observed at shorter wavelengths. The PA of the
object in our VISIR image is compatible with the values found in the
optical and near-IR, but the separation we find at 11.85~$\mu$m is
slightly smaller than that at shorter wavelengths. We conclude
that the elliptical intensity distribution observed by
\cite{Meixner1999} and DWK03 is
due to insufficient spatial resolution in their observations. 

In the deconvolved image, it is apparent that both lobes do not point
at a common centre of symmetry. This was also observed by
\cite{Sahai1999} in the optical HST images. 

Using the central star parameters\footnote{These parameters correspond
to a star at the tip of the AGB, but we can not rule out a more evolved
central star. This would imply a slightly higher temperature for the
small dust grains, but has no major consequences for the results
discussed here.} of DKW03 ($T_{\rm{eff}} = 2670$\,K and
$R_*=372$\,R$_{\odot}$), the distance of 2\,kpc and inclination $ i =
40\degr$ from STM99, and assuming that the dust temperature
$T_{\rm{d}}$ scales with the distance $r$ from the central star as
$T_{\rm{d}} = T_{\rm{eff}}(2r/R_*)^{-0.4}$ \citep[e.g.][]{Herman1986},
we find that $T_{\rm{d}}\,\sim$\, 160\,K at 900\,AU, i.e. the
distance between the central star and the approximate center of either
lobe. Both numbers are uncertain to at least 20\%, but what follows
does not depend so sensitively on these numbers.
Using Wien's law, we find a peak of the energy distribution just to
the red of the N-band. Given the spectral shape of the lobes over the
N-band presented in Sect.\,\ref{sec:spectra}, we conclude that the
N-band image and spectrum are dominated by thermal emission from the
dust in the bipolar lobes, which is heated directly by the central
star. The material in between, which is obscuring the central star,
remains dark even in the MIR, indicating that along the line-of-sight,
optical depth $\tau = 1$ is reached already in the cold ($T < 100$\,K)
outer regions of the dusty environment.

\subsection{The spectra}
\label{sec:spectra}

%
%
\begin{figure}
\centering
  \resizebox{\hsize}{!}{\includegraphics{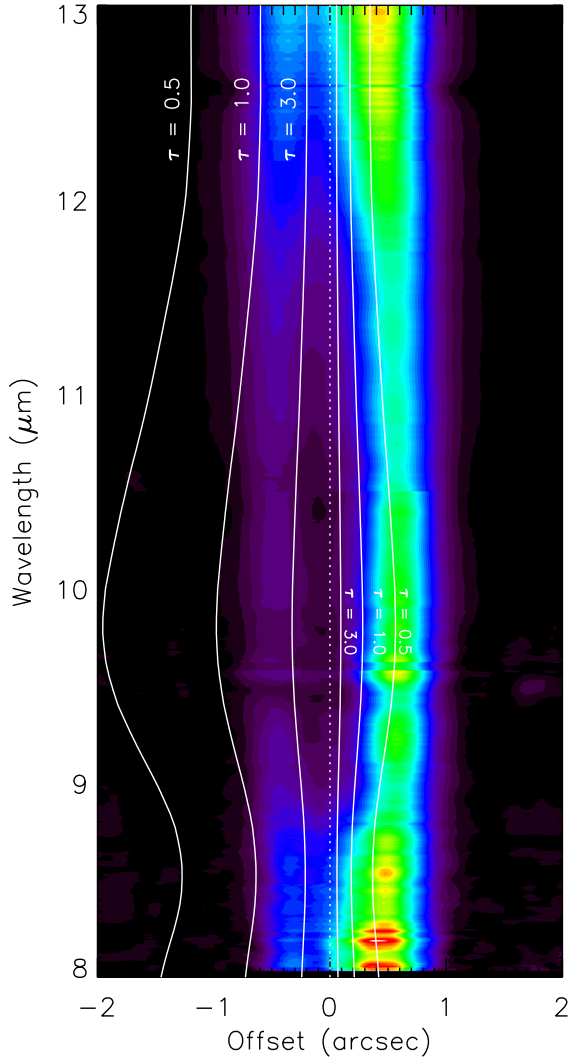}}
  \caption{
 The reduced and calibrated, but not yet collapsed, spectra (4 bands)
 pasted together and normalized to the continuum level as derived from
 a spline fit to the ISO-SWS spectrum (DKW03). The colour scale is
 linear. Perusal of this figure already reveals that the dark waist
 reduces the flux level of both lobes around 10\,$\mu$m, but more so
 for the Eastern lobe. The solid white lines represent contours of
 constant opacity following the model presented in Sect.\,\ref{sec:modelling}. 
}
  \label{fig:2D_spectra}
\end{figure}  

\begin{figure}
\centering
  \resizebox{\hsize}{!}{\includegraphics{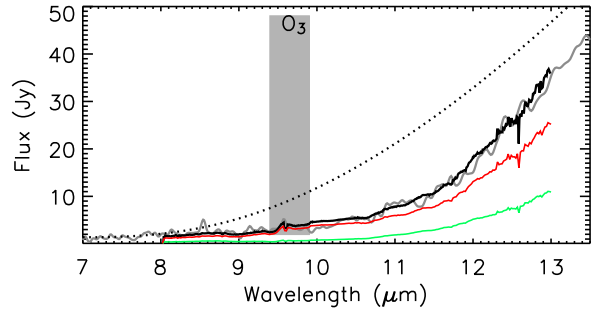}}
  \caption{
 In black, we show the field-integrated pipeline-extracted spectrum calibrated as discussed in
 Sect.~\ref{sec:observations}. For comparison, the ISO-SWS spectrum
 and spline-fit continuum of
 DKW2003 are shown in grey and dotted lines respectively. Shown in red
 and green are the separated spectra of the Western and Eastern lobes
 respectively.  The grey box indicates the wavelength region of strong
 atmospheric ozone absorption. The calibration procedure presented in
 Sect.\,\ref{sec:observations} resulted in reliable data even at
 these wavelengths. 
}
  \label{fig:pipeline_spectrum}
\end{figure}

The four reduced 2D spectra (at 8.5, 9.8, 11.4 and 12.2~$\mu$m) are
shown in Fig.~\ref{fig:2D_spectra}, one above the other following
increasing wavelength. As shown in Fig.\,\ref{fig:sic_image}, slit
losses should at most be of the order of a few percent. From this 2D
spectrum, we can derive the total field-integrated spectrum, shown as
the black solid line in
Fig.~\ref{fig:pipeline_spectrum}, which is of higher quality than the
ISO-SWS spectrum of DKW03 (shown in grey, scaled to match the absolute
IRAS flux level) but has the same global shape and no particular
spectral features\footnote{The absorption line at $12.5-12.6 \mu$m is
  telluric.}: in contrast to what is observed at longer
wavelengths, we don't find a clear indication of crystalline silicates
(e.g. the Forsterite feature at 11.3\,$\mu$m) within the N\,band

%
%

More interesting is the possibility to separate the Eastern and
Western lobes in this 2D spectrum. The slit PA of 55.4$\degr$ is very
close to the object PA of 66$\degr$ derived from the VISIR image
presented in Sect.~\ref{sec:imaging}. To extract position, width and
flux levels for the individual spectra, we performed a row-wise fit
with a double Gaussian intensity distribution. The 2 lobes can be
fully separated through the entire wavelength range.


%
%
\begin{figure}
\centering
  \resizebox{\hsize}{!}{\includegraphics{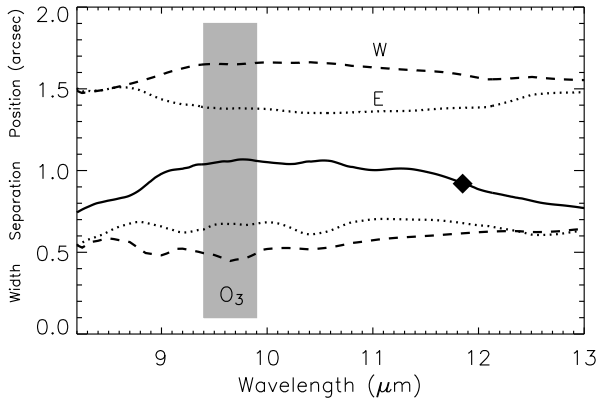}}
  \caption{The dotted and dashed lines represent the position of the E
 and W lobes respectively, minus the position of their continuum at
 8.5~$\mu$m and offset by 1.5~arcsec for clarity.  The solid line
 represents the separation between component E and W as a function of
 wavelength.  The black diamond at 11.85~$\mu$m indicates the
 separation measured in the 2D image presented in
 Sect.~\ref{sec:imaging}. The two curves at the bottom show the width
 of the individual components .
}
  \label{fig:separation}
\end{figure}

The resulting wavelength-dependent positions and widths for both lobes
are shown in Fig.~\ref{fig:separation}. 
The wavelength-dependence of the separation between both lobes clearly
resembles the opacity profile of amorphous silicates, the primary
consituent of the dust grains surrounding the central star. The
presence of such a wavelength dependence shows that the extinction is
not due to a uniform screen of dust (ISM or detached relic of an early-AGB
wind) but instead that, for any given wavelength, the opacity
decreases with distance from some point in between both lobes. This is
also confirmed by the detection of a varying width in the marginally
resolved individual lobes\footnote{The standard star observations show
 a width of $0.27 \pm 0.03$~arcsec.}. Since both lobes are affected by
the dark waist, we find a disagreement with the model presented by
DKW03, in which the flat dusty disk does not result in any obscuration
of the Western lobe.

%
%
\begin{figure}
\centering
  \resizebox{\hsize}{!}{\includegraphics{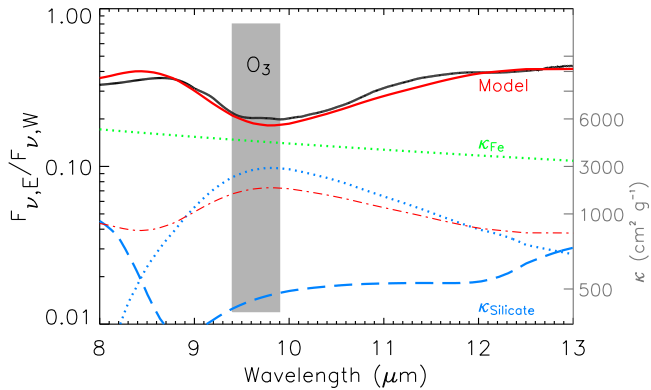}}
  \caption{
  The observed flux ratio between component E and W (black solid line)
 and the model presented in Sect.\,\ref{sec:modelling} (red solid line). Also shown are
 typical absorption (dotted lines) and scattering (dashed lines) cross
 sections for iron grains, and for a mix of small and large silicate
 grains. The scattering by iron grains is negligible at these
 wavelengths. The red dash-dot line represents the $\kappa(\lambda)$
 profile of the best-fit dust composition.
}
  \label{fig:flux_ratio}
\end{figure}  

The individual spectra of the Eastern and Western lobes are presented in
Fig.\,\ref{fig:pipeline_spectrum}, and in Fig.~\ref{fig:flux_ratio} we display
the ratio between both. The individual spectra show that both lobes
emit an equally cold spectrum, and both exhibit the silicate feature in
absorption. This is again in disagreement with the low filling angle
of the model by DKW03 which predicts a dust {\sl emission}
spectrum for the Western lobe.

The flux ratio shows that the Eastern lobe suffers stronger
attenuation than its Western counterpart. Again, the major feature is
that of amorphous silicates.  A proper identification of the type of
silicates would also require the observation of the 18\,$\mu$m band.
The flux ratio ${F_{\nu,\rm{E}}}/{F_{\nu,\rm{W}}} \sim$ 0.4 (instead
of unity) at the ''continuum" wavelengths of 8.5 and 13\,$\mu$m
indicates achromatic attenuation, which is most likely due to metallic
Fe, a crucial ingredient in the SED modelling of OH/IR stars and
obscured RSG \citep{Kemper2002,Harwit2001}.  The observed absorption
towards 8~$\mu$m can be attributed to amorphous silicates, but only
through scattering by fairly large grains. This suggests that a
fraction of the obscuring dust is contained in micron-sized grains,
which was already suggested for the dark waist by DKW03, albeit based
on a geometrically inadequate model as discussed above.

Having found that the obscuring dust structure must have a significant
filling angle, we try to quantify this in the following section.

\section{Interpretation: filling angle}
\label{sec:modelling}

Located at a similar distance as many known Post-AGB binaries, i.e. a
few kpc, IRAS16342 appears much larger in the mid-IR than the stable
dusty disks around those stars: 1\,arcsec vs. 40\,mas
respectively. Equivalently, the observed strong extinction toward the
Eastern lobe requires a significant dust column at least 500\,AU above
the midplane, and this is not possible in a stable Keplerian dust disk
with a reasonable mass. The dust must therefore have some radial velocity.

\begin{figure}
\centering
  \resizebox{\hsize}{!}{\rotatebox{0}{\includegraphics{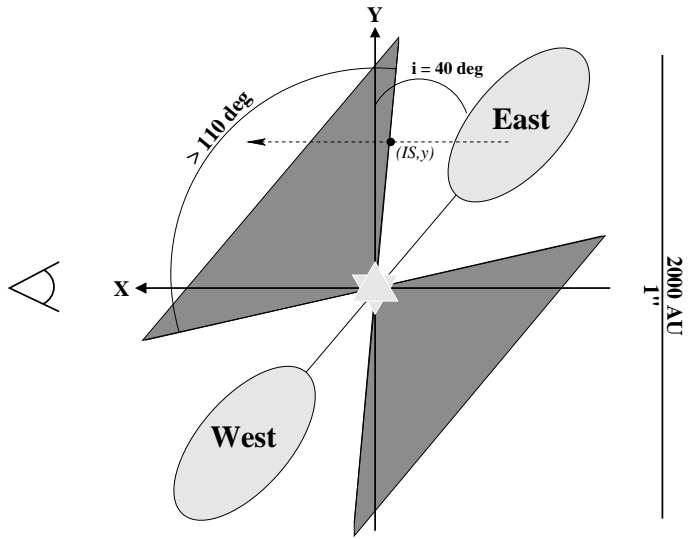}}}
  \caption{The toy model used to estimate the filling angle of the
  obscuring dust structure. The density in the dust structure drops with $r^2$, where $r$
  is the distance to the central star. The inclination is taken from
  SAH99. {\it(IS,y)} is the intersection between a line-of-sight and
  the inner edge of the dust structure. The lobes emit thermal
  radiation at 160\,K.
}
  \label{fig:fillingangle}
\end{figure}

With a simple geometrical model (see Fig.\,\ref{fig:fillingangle}),
which is not intended to be a detailed representation of the source
with all its peculiar characteristics, it
is possible to get an estimate of the filling angle and dust mass of
the obscuring dust structure from the spatially resolved N-band
spectrum.  The mid-IR data
presented here have as advantage over the optical and near-IR data
presented hitherto that the modelling is much less sensitive to
scattering. Including scattering in the analysis would add a new level
of complexity (e.g. the central source spectrum, the scattering
properties of the dust at short wavelengths, the scattering phase
function), and is not required to constrain the basic structural
parameters. We leave the construction of a full "colour" image to a
future paper. 

%
%

In the assumption of an outflow with a biconical cavity, the optical
depth $\tau$ along a line-of-sight at a distance $y$ from the center
is calculated as:
\begin{eqnarray}
 \tau (y,\lambda)_{E,W} & = & \int_{IS}^{\infty}\frac{\rho_0\kappa(\lambda)}{x^2+y^2}dx \nonumber\\
& = & \frac{\rho_0\kappa(\lambda)}{y^2}[\arctan(x/y)]^{\infty}_{y\tan(\frac{\pi}{2}-\frac{\theta}{2}\pm
i)} \nonumber\\
& = & \frac{\rho_0\kappa(\lambda)}{y^2}(\frac{\theta}{2}\pm i).\nonumber
\end{eqnarray}
It depends on (1) the filling angle $\theta$ of the dust structure,
(2) the dust density $\rho_0$ at unit radius\footnote{The density
follows a r$^{-2}$ distribution, with a 15\,km\,s$^{-1}$ outflow
velocity.}, and (3) the dust mass fraction caught in small and large
silicate grains\footnote{We use only these 2 discrete sizes since
there is no reason to believe that a classical power law size
distribution is valid for this dust structure.} (0.1 and 5\,$\mu$m),
and in metallic iron, implicit in the wavelength-dependent cross
section $\kappa$. The "background" radiation by the lobes is assumed
to be thermal emission by small dust grains at 160\,K, as derived in
Sect.\,\ref{sec:imaging}. Whether these small dust grains fill the
cavities or form the boundary with the denser dust structure cannot be
derived from our current observations. We do not assume an actual
density or temperature distribution for the individual lobes: the
lobes appear marginally resolved in our observations, but
insufficiently so to derive such a distribution. The comparison
between model and observations is therefor done at the "nominal"
distances of $y_0=0.45$\,arcsec (900\,AU) from the center of the
object.  We determine the unknowns by requiring reproduction of (1)
the very red SED up to 20\,$\mu$m with total obscuration of the
central star and hot inner dust (from the optical to the mid-IR), (2)
the wavelength-dependent flux ratio between both lobes, and (3) the
moderate observed extinction towards the Western lobe
($\tau_{10\mu\rm{m}} \sim 0.5$, derived from the depth of the silicate
absorption feature in the Western spectrum). The wavelength dependence
of the dust opacity is mostly constrained by (2). The dust density is
determined by (1) and (3).

We find that a filling angle of at least 110\,$\deg$ is required to
produce observable extinction towards the Western lobe. The best fit
is obtained with a 145\,$\deg$ filling angle. With a dust species
mass ratio of $\frac{\rho_{silicates}}{\rho_{Fe}}=5.5$, a silicate size
dust mass ratio of $\frac{\rho_{0.1\mu m}}{\rho_{5.0\mu
    m}}=1.5$, and a
mass-loss rate of $\dot{M} \sim 1 \times
10^{-3}$\,M$_{\odot}$\,yr$^{-1}$, we find good agreement with the
N-band observations (Fig.\,\ref{fig:flux_ratio}) and lack of flux at
shorter wavelengths (toward the central star, our model predicts $\tau
\ge 100$ up to 30$\mu$m). However, we can not exclude an even larger
filling angle.

\section{Discussion and conclusions}
\label{sec:conclusions}

We find that, in spite of previous classifications as an elliptical
MIR source, IRAS16342 in fact appears as a bipolar source up to at
least 13\,$\mu$m, if studied with sufficient angular
resolution. Almost all of the N-band flux originates in these 2 lobes,
with an apparant temperature of roughly 160\,K. The dusty ``waist''
region remains dark even at these wavelengths. From the varying
extinction in front of both lobes as a function of wavelength, we
derive that the dust structure has a large filling angle
($\sim$145$\degr$), much larger than the stable Keplerian dust disks
recently seen around other binary Post-AGB stars \citep[typically
40$\degr$, ][]{deruyter2006,Deroo2007}, and this in spite of the strong indications
that also IRAS16342 has a binary central system. Instead, it resembles
more the spherically symmetric superwind of an OH/IR star, but with
cavities blown by jets.  Unfortunately, the dam around this fountain
is too high to see the engine driving the bipolar jets. It will be
interesting to see whether all members of the class of Water Fountain
Nebulae display this kind of dust structure.

%

\acknowledgements{The authors would like to thank the anonymous
  referee for many valuable comments, and E.\,Lagadec for a careful
  reading of the manuscript.}

\bibliographystyle{aa}

\end{document}